\documentclass[twocolumn]{aastex631}
\usepackage{amsmath}
\usepackage{mathrsfs}

\graphicspath{{fig/}}
\defcitealias{Tominaga2021}{TIK21}
\usepackage[T1]{fontenc}

\begin{document}
\title{The Dynamical Interaction between Low-mass Planets and Dust Coagulation}
\correspondingauthor{Cong Yu}
\email{yucong@mail.sysu.edu.cn}
\author[0000-0002-8125-7320]{Qiang Hou}
\affiliation{School of Physics and Astronomy, Sun Yat-sen University, Zhuhai 519082, China}
\affiliation{CSST Science Center for the Guangdong-Hong Kong-Macau Greater Bay Area, Zhuhai 519082, China}
\affiliation{State Key Laboratory of Lunar and Planetary Sciences, Macau University of Science and Technology, Macau, China}
\author[0000-0003-0454-7890]{Cong Yu}
\affiliation{School of Physics and Astronomy, Sun Yat-sen University, Zhuhai 519082, China}
\affiliation{CSST Science Center for the Guangdong-Hong Kong-Macau Greater Bay Area, Zhuhai 519082, China}
\affiliation{State Key Laboratory of Lunar and Planetary Sciences, Macau University of Science and Technology, Macau, China}
\affiliation{International Centre of Supernovae, Yunnan Key Laboratory, Kunming 650216, China}
\author[0000-0003-4366-6518]{Shu-ichiro Inutsuka}
\affiliation{Department of Physics, Graduate School of Science, Nagoya University, Furo-cho, Chikusa-ku, Nagoya, Aichi 464-8602, Japan}

\begin{abstract}
We investigate the impact of a low-mass planet on dust coagulation, and its consequent feedback on planetary migration, using a linear analysis of the coupled dust-gas hydrodynamic equations. Dust coagulation is incorporated via a single-size approximation. In the co-orbital region of the planet, we find that the growth of dust size is significantly suppressed by planet-induced coagulation modes (CMs). This effect are less pronounced with smaller stopping times, stronger gaseous turbulence or imperfect sticking. Regarding planetary migration, we find that CMs make outward migration require $\tau \gtrsim 0.3$ ($\tau$ is dimensionless stopping time) with typical turbulent strength and dust coagulation efficiency. We demonstrate that the torque variations are reasonable and arise from phase shifts between the density and stopping time perturbations in the coagulation modes.
\end{abstract}

\keywords{Hydrodynamics (1963), Protoplanetary disks (1300), Planetary-disk interactions (2204), Planetary migration (2206), Planet formation (1241)}

\section{Introduction} \label{introduction}
Dust coagulation is a fundamental process in protoplanetary disks (PPDs), driving the growth of sub-micron grains into larger aggregates and ultimately leading to planetesimal formation \citep{Blum2008, Testi2014, Birnstiel2016}. While the presence of planets alters dust evolution by creating pressure bumps, gaps, and vortices, which can either enhance or suppress coagulation \citep{Zhu2012,Pinilla2012,Weber2018}. Studies have shown that Jupiter-mass planets trap dust in pressure bumps, increasing local dust-to-gas ratios and facilitating planetesimal formation \citep{Joanna2019}. Moreover, planetary perturbations can regulate dust size distributions by influencing collisional velocities and coagulation efficiency \citep{Eriksson2024}. While these effects have been studied for massive planets, the impact of low-mass planets (e.g., Earth-mass planets) on dust coagulation remains largely unexplored. Investigating the impact is critical for understanding the early stages of planet formation when growing protoplanets are still accreting dust grains.

Theoretically, dust coagulation is described by the Smoluchowski equation \citep{Smoluchowski1916}. Recent advancements have allowed the coupling of coagulation models with hydrodynamic simulations, providing a more comprehensive view of dust evolution \citep{Joanna2019,Laune2020,Li2020,Ho2024}. However, solving the full Smoluchowski equation is computationally expensive, particularly when coupled with disk dynamics. Alternatively, a single-size approximation has been developed, offering an efficient way to describe dust coagulation \citep{Ormel2008,Estrada2008,Sato2016}. In this study, we adopt the approximation and perform linear analysis to explore the interaction between low-mass planets and dust coagulation. Our work offers new insights into the coupling between planets and dust growth.

The paper is organized as follows. In \autoref{method}, we present the basic equations. In \autoref{results}, we present our results. In \autoref{conclusions}, we present the discussions and conclusions.

\section{Methods} \label{method}
\subsection{Two-fluid Equations}
We assume the presence of a low-mass planet with mass $M_p$, embedded in a dust-gas-coupled PPD with surface density $\Sigma$. The planet orbits a star of mass $M_{\star}$ at an angular frequency $\Omega_p$ with orbital radius $r_p$ and azimuthal angle $\theta_p$. The gas is assumed isothermal. The dust is treated as pressureless fluid and diffused by gaseous turbulence. The strength of turbulence is described by $\alpha$ \citep{Shakura1973}. Then the advected velocities of dust can be decomposed into the mean flow velocities and diffusion velocities, i.e. $\boldsymbol{u}_d^{*} = \boldsymbol{u}_d + \boldsymbol{u}_{d,\rm{diff}} $ and $\boldsymbol{u}_{d,\rm{diff}} = - D \nabla \left( \ln \mu \right)$. $\mu = \Sigma_d / \Sigma_g$ is the dust-to-gas mass ratio and $D = \alpha c_s H$ is the diffusion coefficient, where $c_s$ is the sound speed and $H$ is the disk scale height. In the following, we will use the advected velocities instead of the mean flow velocities for dust. And for convenience, we drop the superscript ``$*$'' for ``$u_d^{*}$''. Then, the set of equations governing the dynamics of dust and gas within the shearing-sheet approximation \citep[e.g.,][]{Goldreich1965}, which defines a Cartesian coordinate: $x = r - r_p, y = r_p(\theta-\theta_p)$, is given by
\begin{equation} \label{eq1}
    \frac{\partial \Sigma_g}{\partial t}+\nabla \cdot\left(\Sigma_g \boldsymbol{u}_g\right) =0 \ ,
\end{equation} 
\[
    \frac{\partial \boldsymbol{u}_g}{\partial t}+\boldsymbol{u}_g \cdot \nabla \boldsymbol{u}_g= - 2 \boldsymbol{\Omega}_p \times \boldsymbol{u}_{g} - \nabla \left( \Phi_{p} + \Phi_{\rm{tid}} \right) \nonumber 
\]
\begin{equation} \label{eq2}
    + \frac{\mu \boldsymbol{w}_s}{t_{\text {s}}}-c_s^2 \nabla \ln \Sigma_g + 2\eta \Omega_p^2 r_p \boldsymbol{e}_{x} \ ,
\end{equation}
\begin{equation} \label{eq3}
    \frac{\partial \Sigma_d}{\partial t}+\nabla \cdot\left(\Sigma_d \boldsymbol{u}_d\right) = 0 \ ,
\end{equation}
\[
    \frac{\partial \boldsymbol{u}_d}{\partial t}+\boldsymbol{u}_d \cdot \nabla \boldsymbol{u}_d= - 2 \boldsymbol{\Omega}_p \times \boldsymbol{u}_{d} - \nabla \left( \Phi_{p} + \Phi_{\rm{tid}} \right) \nonumber 
\]
\begin{equation} \label{eq4}
    -\frac{\boldsymbol{w}_s}{t_{\text {s}}} - \frac{D}{t_s} \nabla \ln \mu .
\end{equation}
Here we use the conventional notation for the quantities. And $\eta = - 1/(2\Sigma V_{\mathrm{K}}^2) \partial P / \partial \ln r $ is the global pressure \citep{Youdin2005}. The planetary potential $\Phi_p$ scales linearly with $M_p$. $\Phi_{\rm{tid}} = -1.5 \Omega_p (r - r_p)^2$ is the tidal potential. $\boldsymbol{w}_s = \boldsymbol{u}_d - \boldsymbol{u}_g$ represents the relative velocity between dust and gas. The stopping time, $t_s$, characterizes the coupling strength between dust and gas. We adopt its dimensionless form, $\tau = \Omega_p t_s$, in the following discussions. In most cases, it is identical to the Stokes number.

\subsection{Dust Coagulation Equation}
We model dust coagulation using the single-size approximation, which has been validated in Appendix A of \citet{Sato2016}. Specifically, we assume the dust mass distribution at $r_p$ is narrowly peaked at a representative mass $m_p$ that dominates the surface density $\Sigma$. The evolution of $m_p$ is governed by
\begin{equation}
\frac{d m_{p}}{d t}= \varepsilon_{\mathrm{eff}} \frac{2 \sqrt{\pi} a^2 \Delta v_{\mathrm{pp}}}{H_{\mathrm{d}}} \Sigma_{\mathrm{d}}, \label{eq_coa}
\end{equation}
where $a$ is the grain size, $H_d \simeq \sqrt{\alpha/\tau} H$ is the dust scale height, and $\Delta v_{\mathrm{pp}}$ is the relative velocity between colliding grains. We consider only turbulence-driven collisions.

In the single-size approach, the best agreement with full Smoluchowski simulations is achieved when the colliding particles have stopping times $\tau_1 = \tau(m_p)$ and $\tau_2 = 0.5\,\tau(m_p)$ \citep{Sato2016}. Throughout this work, we adopt this configuration and restrict our analysis to intermediate dust sizes with $\mathrm{Re}^{0.5} < \tau < 1.0$, where $\mathrm{Re}$ is the Reynolds number. In this regime, the relative velocity is approximated as $\Delta v_{\mathrm{pp}} \simeq \sqrt{2.3\,\tau\alpha}\, c_s$ \citep{Ormel2007}.

When $\Delta v_{\mathrm{pp}}$ is small, collisions lead to perfect sticking and efficient coagulation. However, at higher velocities, additional processes such as bouncing or erosion can occur, reducing the net growth efficiency and potentially leading to fragmentation \citep{Guttler2010,Krijt2015}. To capture these effects, we introduce a sticking efficiency $\varepsilon_{\mathrm{eff}}$ \citep{Wada2009,Wada2013,Okuzumi2016,Ueda2019}, given by
\begin{equation} \label{eq_eff}
\varepsilon_{\mathrm{eff}} = \min \left(1,-\frac{\ln \left(\Delta v_{\mathrm{pp}} / v_{\mathrm{frag}}\right)}{\ln 5}\right),
\end{equation}
where $v_{\mathrm{frag}}$ is the fragmentation threshold. For $\Delta v_{\mathrm{pp}} \lesssim v_{\mathrm{frag}}$, we have $0 < \varepsilon_{\mathrm{eff}} < 1$, corresponding to imperfect sticking. When $\Delta v_{\mathrm{pp}} > v_{\mathrm{frag}}$, $\varepsilon_{\mathrm{eff}} < 0$, indicating net fragmentation.

We further assume that dust grains are spherical (then $m_p = 4/3 \pi \rho_{\mathrm{int}} a^3$) and reside in the Epstein drag regime, where the stopping time is given by
\begin{equation}
\tau=\frac{\pi}{2} \frac{\rho_{\mathrm{int}} a}{\Sigma_{\mathrm{g}}}, \label{eq_Epstein}
\end{equation}
with $\rho_{\mathrm{int}}$ being the internal density of a grain.

Substituting \autoref{eq_Epstein} into \autoref{eq_coa}, we obtain the evolution equation for the stopping time:
\begin{equation} \label{eq5}
    \frac{\partial \ln t_s}{\partial t}+ \boldsymbol{u}_{d} \cdot \nabla \ln t_s = \varepsilon_{\mathrm{eff}} \frac{\mu}{3 t_0} + \nabla \cdot \boldsymbol{u}_g - \boldsymbol{w}_s \cdot \nabla \ln \Sigma_g,
\end{equation}
where $t_0 \equiv 4/3 \sqrt{2.3 \pi } \Omega_p^{-1} \simeq 0.49 \Omega_p^{-1}$ is the characteristic mass growth timescale of a dust grain. And the timescale of dust coagulation is $3t_0/\mu$.

On the right-hand side of \autoref{eq5}, the first term represents growth due to coagulation. The second term accounts for gas compression. And a negative value indicates convergence of gas flow, which reduces $\tau$. The third term reflects the change in stopping time as dust drifts toward regions of higher gas surface density. However, this effect is negligible in our case due to the weak gas density gradient.

\subsection{Linear Analysis}
We adopt linear analysis. Without planetary potential, the equilibrium state follows the Nakagawa-Sekiya-Hayashi (NSH) equilibrium \citep{Nakagawa1986}. Strictly speaking, this is a quasi-equilibrium state because of dust coagulation. It requires that the coagulation timescale is much longer than the dynamical timescale, i.e., $3 t_0 / \mu \gg \Omega_p^{-1}$. To satisfy this condition, the dust-to-gas mass ratio should be sufficiently small, specifically $\mu \ll 1.5$. In this study, we adopt $\mu = 1/99$ unless stated otherwise.

We then assume that the first-order perturbation of any physical quantity can be expressed as $X_1 = \int d k_y \delta X \exp \left( ik_y y  \right)$, where $k_y$ is the azimuthal wavenumber. The linearized perturbation form of \autoref{eq1}-\ref{eq4} and \autoref{eq5} are presented in \autoref{appendix}. This approach is valid when the planetary mass is sufficiently small, specifically when $M_p \ll M_{\mathrm{th}}$. The thermal mass $M_{\mathrm{th}} = h_p^3 M_{\star}$, where $h_p = H / r_p$ is the disk aspect ratio, which is set to 0.03 in this study. It represents the threshold above which the planet significantly perturbs the disk structure.

Finally, because of \autoref{eq_Epstein}, there exists a relationship between $\delta a$ and $\delta \tau$, given by
\begin{equation}
    \frac{\delta a }{a} = \frac{\delta \Sigma_g }{ \Sigma_g} + \frac{ \delta \tau}{\tau},
    \label{eq_dust_size}
\end{equation}
where the quantities in the denominators represent their respective equilibrium values.

\section{Results} \label{results}
In this section, we solve \autoref{linear1}--\ref{linear_coa} using the numerical method described in \citet{Hou2025}, and examine the impact of a planet on dust coagulation, along with the back-reaction of coagulation on the planetary migration. Unless stated otherwise, the sticking efficiency $\varepsilon_{\mathrm{eff}}$ is assumed to be $1$, corresponding to perfect sticking. The approach is formally valid when $M_p \ll M_{\mathrm{th}}$. Nevertheless, for convenience we adopt $M_p = M_{\mathrm{th}}$ in the calculations. This does not affect the results, because the problem is linear; the choice of a large $M_p$ is solely for illustrative purposes, except when the results are normalized.

\subsection{Coagulation Mode} \label{mode}
\begin{figure}[htbp!]
    \centering
    \includegraphics[width=1.0\columnwidth]{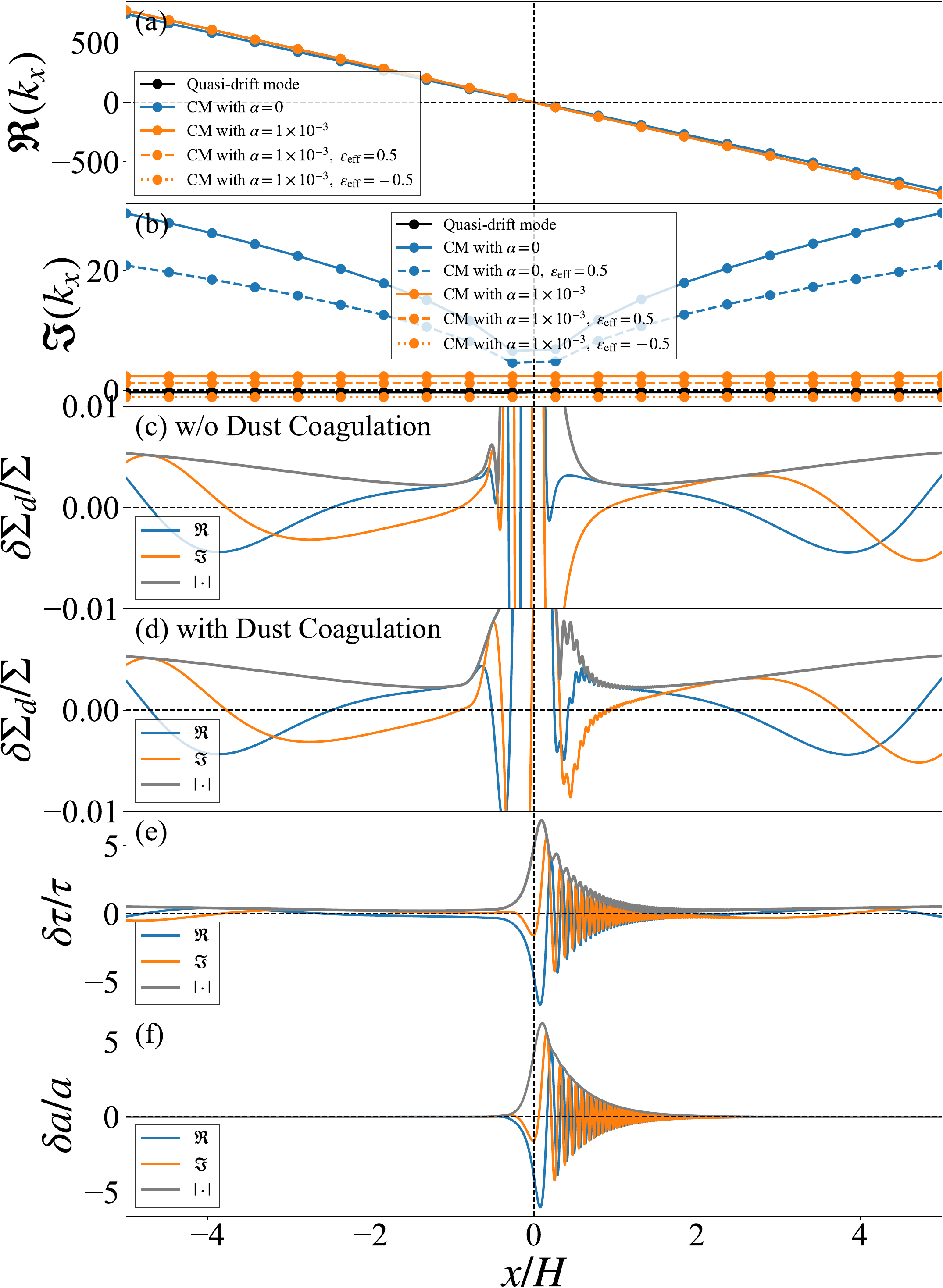}
    \caption{Panel (a) and (b): The real and imaginary part of $k_x$ for the quasi-drift mode and CMs. The solid dots represent our numerical results. Panel (c)-(f): the wave functions normalized by their respective equilibrium values. The parameters are set as $\tau = 0.1$, $k_y = 0.3$ and $\alpha = 1\times 10^{-3}$. In those four panels, the blue, orange and gray lines represent the real part, imaginary part and amplitude, respectively. } 
    \label{fig1}
\end{figure}
Without the planetary potential, \autoref{linear1}--\ref{linear_coa} own seven eigenmodes, six of which have been analyzed in \cite{Hou2025}. These include gaseous and dusty density waves (DWs), and the quasi-drift modes. The seventh mode, introduced by dust coagulation, is referred to as the ``coagulation mode (CM)'', which can give rise to coagulation instability (CI) \citep[][TIK21 hereafter]{Tominaga2021}. The CM is a mode that leads to variations in the stopping time (or equivalently, dust size). \citetalias{Tominaga2021} derived a dispersion relation of CMs:
\begin{equation}
\begin{aligned}
\tilde{\omega} = & k_x u_{dx} \\
& + i \varepsilon_{\mathrm{eff}} \frac{\mu}{6 t_0}\left(1 + \sqrt{1-\varepsilon_{\mathrm{eff}}^{-1} \frac{12 t_0}{\mu} \frac{1-\tau^2}{1+\tau^2} i k_x u_{dx}}\right),
\end{aligned}
\end{equation}
where $\omega$ and $\tilde{\omega} \equiv \omega - k_y h_p^{-1} \Omega_p$ are the eigenfrequency, Doppler-shifted frequency, respectively. We note that the oscillation rate $\Re (\tilde{\omega}) \simeq k_x u_{dx}$, which is similar to that of the quasi-drift mode \citep{Hopkins2018,Hou2024}, which means CMs are also induced by dust drift velocity. And the growth rate $\Im (\tilde{\omega})$ is proportional to $\sqrt{\varepsilon_{\mathrm{eff}}}$ at short wavelength.

In our forced oscillation set-up, $\tilde{\omega}$ is real, while the radial wavenumber $k_x$ is complex. The growth rate corresponds to the positive value of the imaginary part of $k_{x}$, both of which imply that the waves are growing while propagating. Panel (a) in \autoref{fig1} presents $\Re(k_x)$ for the quasi-drift mode and CM with $k_y = 0.3$ and $\tau = 0.1$. It is evident that $\Re(k_x)$ for CMs closely matches that of the quasi-drift mode and remains unchanged for different values of $\alpha$ and $\varepsilon_{\mathrm{eff}}$, demonstrating the above statement. With $\alpha=0$, Panel (b) shows that the ratio of $\Im(k_x)$ between $\varepsilon_{\mathrm{eff}}=0.5$ and $\varepsilon_{\mathrm{eff}}=1$ is $\sqrt{\varepsilon_{\mathrm{eff}}}$. But with a non-zero $\alpha$, the ratio changes to 0.5. At the same time, turbulent diffusion will decrease $\Im(k_x)$ significantly. Anyway, all $\Im(k_x)$ of them are positive, while those of the quasi-drift mode and CM with $\varepsilon_{\mathrm{eff}}=-0.5$ are negative, implying they are stable modes in an eigenvalue problem.

With the planetary potential, the wave functions normalized by their respective equilibrium values are presented in Panel (c)--(f). The $x$-axis represents radial direction. Panels (c) and (d) show the dust density perturbations $\delta \Sigma_d$ without and with dust coagulation, respectively. Panel (d) reveals that $\delta \Sigma_d$ exhibits fine structures outside the planet compared to panel (c), which are induced by the CM. Panel (e) presents the stopping time perturbations $\delta \tau$, demonstrating that the CM is mainly present outside the planet, characterized by an extremely short wavelength. Panel (f) shows the dust size perturbations $\delta a$. Notably, the long-wavelength component observed in $\delta \tau$ disappears in $\delta a$. According to \autoref{eq_dust_size}, this implies that $\delta \tau$ is primarily influenced by gas compression and CM, while the long-wavelength component originates from gaseous DWs.

\subsection{Revisiting of Coagulation Instability} \label{mechanism}
\begin{figure*}[htbp]
    \centering
    \includegraphics[width=1.0\textwidth]{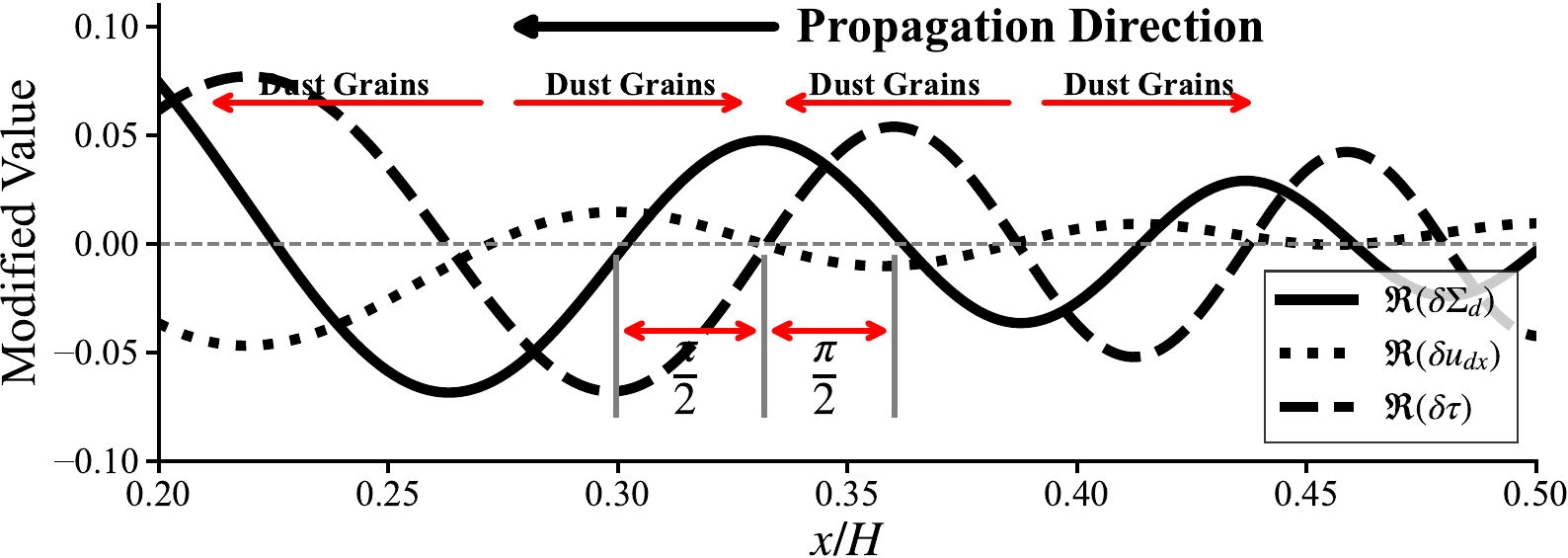}  
    \caption{The waveforms of CM, which illustrates the mechanism of CI. The three waveforms represent the real parts of the perturbations in dust surface density $\delta \Sigma_d$, radial velocity $\delta u_{dx}$, and stopping time $\delta \tau$.  Their amplitudes have been rescaled to same order. The propagation direction of the waves is inward, as indicated by the black arrow. A positive $\delta \Sigma_d$ leads to an increase in $\delta \tau$ with a phase shift of $\pi/2$ due to dust coagulation. As $\delta \tau$ increases, the aerodynamic drag exerted by the gas on the dust decreases, which in turn increases the dust radial velocity. In the figure, $\delta u_{dx}$ is approximately out of phase with $\delta \tau$. Consequently, the phase difference between $\delta u_{dx}$ and $\delta \Sigma_d$ is close to $\pi/2$, implying that dust grains tend to accumulate near the crests of $\delta \Sigma_d$ and disperse from the troughs, as indicated by the red arrows. This leads to the amplification of the initial $\delta \Sigma_d$. The positive feedback among $\delta \Sigma_d$, $\delta \tau$, and $\delta u_{dx}$ ultimately drives the CI.}
    \label{fig2}
\end{figure*}

\citetalias{Tominaga2021} interpreted CI as a positive feedback loop between dust surface density and stopping time. Here, we incorporate the role of the radial velocity and revisit its mechanism. In the absence of dust diffusion, \citetalias{Tominaga2021} found, through analytical calculations, that the phase shift between the dust density perturbation $\delta \Sigma_d$ and the stopping time perturbation $\delta \tau$ is $\pi/4$, while the phase shift between the radial velocity perturbation $\delta u_{dx}$ and the dust density perturbation is $3\pi/4$. However, for CMs with longer wavelengths, both of the two phase shifts will become $\pi/2$ \citep{Tominaga2022a,Tominaga2022b}. We find that when dust diffusion is non-zero, the phase shifts approach $\pi/2$ for all CMs. This is because dust diffusion lengthens the wavelengths. And the critical turbulent efficiency for the shift is about $\alpha = 1 \times 10^{-5}$, which is easy to be achieved. Therefore, we will consider a $\pi/2$ phase shift between $\delta u_{dx}$, $\delta \Sigma_d$ and $\delta \tau$ hereafter, which is also helpful to present a clear physical interpretation for CI mechanism.

We present a schematic illustration in \autoref{fig2} for the case with $k_y = 0.3$, $\tau = 0.1$, and $\alpha = 1 \times 10^{-4}$. The figure shows three waveforms, corresponding to the real parts of $\delta \Sigma_d$, $\delta \tau$ and $\delta u_{dx}$. These waveforms represent the differences between simulations with and without dust coagulation, and thus characterize the CMs. To facilitate the comparison, the amplitudes of the waveforms are rescaled to comparable levels. The waves propagate inward, as indicated by the black arrow.

As shown in the figure, $\delta \Sigma_d$ leads $\delta \tau$ by a phase shift of $\pi/2$, implying that positive $\delta \Sigma_d$ induces an increase in $\delta \tau$. Notably, the ratio of the amplitudes of $\delta \tau$ and $\delta \Sigma_d$ is $\mu / 3t_0$ which shows the direct effect of the coagulation term in \autoref{eq5}. An increase in $\delta \tau$ reduces the aerodynamic drag exerted by the gas on the dust, which in turn increases $\delta u_{dx}$. In the figure, $\delta u_{dx}$ is out of phase with $\delta \tau$. Consequently, the phase shift between $\delta u_{dx}$ and $\delta \Sigma_d$ is approximately $\pi/2$, meaning that dust grains tend to accumulate near the crests of $\delta \Sigma_d$ and disperse from the troughs, as illustrated by the red arrows. This behavior leads to an amplification of the initial $\delta \Sigma_d$. The positive feedback among $\delta \Sigma_d$, $\delta \tau$, and $\delta u_{dx}$ ultimately gives rise to the CI.

\subsection{Suppressed Growth of Dust Size} \label{dust_size}
\begin{figure*}[htbp]
    \centering
    \includegraphics[width=1.0\textwidth]{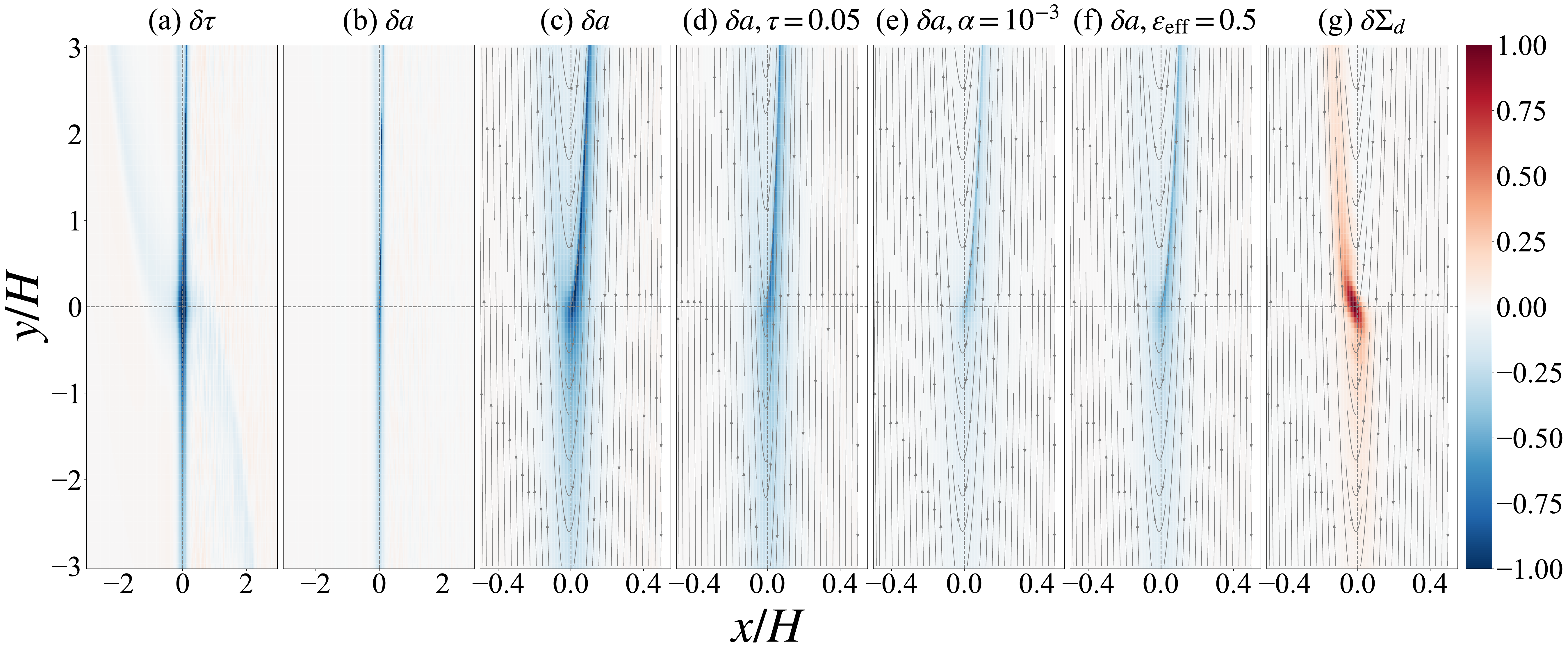}  
    \caption{2D morphologies illustrating the effects of a planet on dust coagulation, with $\tau = 0.1$, $\alpha = 1 \times 10^{-4}$ and $\varepsilon_{\mathrm{eff}}=1.0$ except for Panel (d), (e) and (f). The $x$- and $y$-axes correspond to the radial and azimuthal directions, respectively. Subtitles within each panel indicate the specific parameters used. Panel (a) show the perturbations in stopping time. Panel (b) shows the perturbations in dust size. Panels (c) and (d) provide the zoomed-in views of the dust size perturbations for $\tau = 0.1$ and $\tau = 0.05$, respectively. Panel (e) shows the dust size perturbations for $\alpha = 1 \times 10^{-3}$. Panel (f) shows the dust size perturbations with $\varepsilon_{\mathrm{eff}}=0.5$. Panel (g) shows the dust density perturbations with $\varepsilon_{\mathrm{eff}}=0.5$. Gray solid lines with arrows those Panels indicate the equilibrium streamlines. The colorbar denotes the normalized values. For a comparison, the values in Panel (b)-(f) are normalized by their common maximum value.}
    \label{fig3}
\end{figure*}
The results for the perturbations in stopping time, dust size and density are presented in \autoref{fig3}, with $\tau = 0.1$, $\alpha = 1 \times 10^{-4}$ and $\varepsilon_{\mathrm{eff}}=1.0$ except for Panel (d), (e) and (f). The figure consists of seven panels, each labeled with a subtitle. For instance, Panel (a) displays $\delta \tau$ in the case with $\tau = 0.1$, $\alpha = 1 \times 10^{-4}$ and $\varepsilon_{\mathrm{eff}}=1.0$. Although we have performed calculations over a wide range of $\tau$ (specifically, $ 0.01\leq \tau \leq 1.0$), we present only the results for $\tau = 0.1$ for an illustration, as the subsequent findings remain valid across the entire range. The values in Panel (a) are normalized by its one quarter of maximum value in order to make the color more obvious. The values in Panel (b)-(f) are normalized by their common maximum value for a comparison.

From Panel (a), we can see that $\delta \tau$ is positive in in most regions (although the increase is not pronounced), because of dust coagulation, while $\delta \tau$ is negative in two regions. One of these corresponds to the region where gaseous DWs are propagating. But the negative $\delta \tau$ there is stationary and does not mean that gas compression limit the growth of $\tau$ (or $a$). Using \autoref{eq_dust_size}, we can derive the perturbations in dust size, as shown in Panel (b). It is evident that there is no decrease in dust size at the locations of gaseous DWs, which aligns with the previous discussion and the results presented in \autoref{fig1}.

The second region of decrease for both $\tau$ and $a$ corresponds to the co-orbital region of the planet. In this region, $\delta \tau$ and $\delta a$ are nearly identical, so we present a zoomed-in view of $\delta a$ in later panels. Due to the similarity between CMs and the quasi-drift mode, $\delta a$ exhibits a pattern similar to that of the dusty wake along the streamlines given by the NSH velocity \citep[Panel (g) and][]{Hou2025}, as delineated by the gray solid lines with arrows. The dusty wake is composed of many quasi-drift modes, and similarly, the $\delta a$ wake consists of many CMs. However, the key difference is that the dusty density wake has a positive value and is primarily located inside the planet as Panel (g) shows, whereas the $\delta a$ wake has a negative value and is predominantly found outside the planet (where $x$ is positive). Here we should note that the sizes of all dust grains are increasing due to our equilibrium state. And the negative $\delta a$ does not imply that the dust size decreases, but grows more slowly than that in the equilibrium state, i.e. the dust growth is suppressed near in the co-orbital region of the planet.

The physical mechanism by which dust growth is suppressed by planet-induced CMs can be understood as follows. Planets tend to attract dust grains, and as a result, the aerodynamic drag from the gas acts in the outward direction to counterbalance this inward drift in the region outside the planet ($+x$ region). This leads to a negative variation in the stopping time. Dust grains traveling along equilibrium streamlines that intersect the planet are most strongly influenced, giving rise to a distinct wake of reduced stopping time leading the planet. The suppression becomes more pronounced closer to the planet due to stronger interactions, resulting in suppressed dust growth in the co-orbital region. Inside the planet, the mechanism of the interaction force mentioned above becomes opposite, which leads to the asymmetry.

By comparing Panel (c) and (d), we show that the suppression effect is more significant for larger $\tau$ (at least when $0.01 \leq \tau \leq 1.0$), which is consistent with weaker dust-gas coupling at larger stopping times. Panel (e) presents the case with $\tau = 0.1$ and $\alpha = 1 \times 10^{-3}$, showing that increased dust diffusion mitigates the influence of planets on dust growth. This reduction occurs because diffusion smooths out dust density perturbations, thereby weakening the CMs responsible for suppressing coagulation in the co-orbital region. Panel (f) shows the case with $\tau = 0.1$ and imperfect sticking, specifically, $\varepsilon_{\mathrm{eff}}=0.5$. And we found the imperfect sticking makes the effect weaker through comparing it with Panel (c).

Panel (g) presents the dusty density wake with perfect dust coagulation, which can lead to streaming torque \citep[ST,][]{Hou2024,Hou2025}. From this image, we can hardly distinguish the density variations caused by dust coagulation from the previous calculations. In the next subsection, we will focus on that and calculate the disk torque.

\subsection{The Effects of Coagulation Modes on Streaming Torque} \label{sec_torque}
\begin{figure}[htbp!]
    \centering
    \includegraphics[width=1.0\columnwidth]{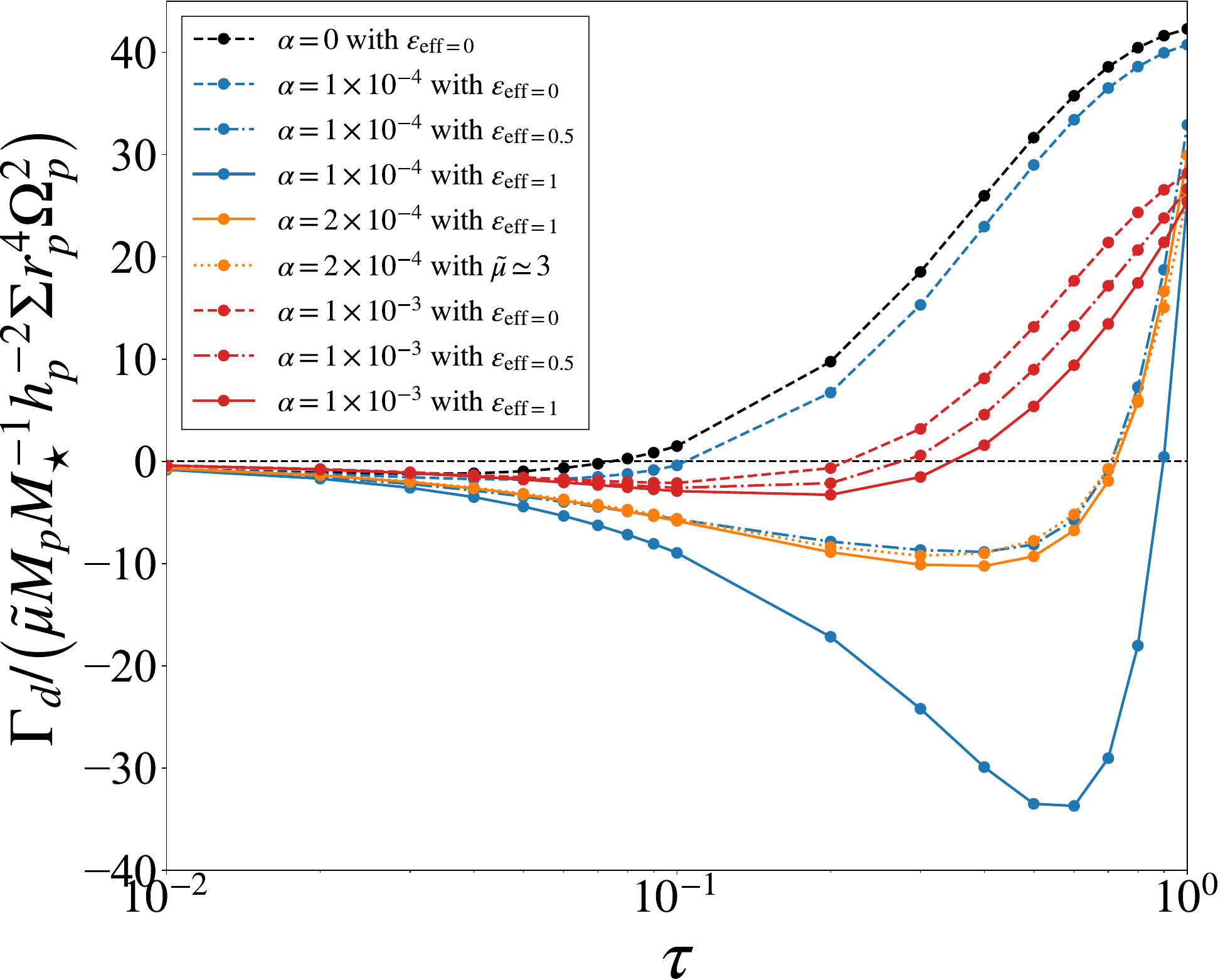}
    \caption{Dusty torque on the planet with different values of $\varepsilon_{\mathrm{eff}}$ and $\alpha$. Lines with different colors correspond to different $\alpha$ as the legend labels. Dashed lines represent the results with $\varepsilon_{\mathrm{eff}} = 0$, i.e., without dust coagulation. Dot-dashed lines represent the results with $\varepsilon_{\mathrm{eff}} = 0.5$. Solid lines represent the results with $\varepsilon_{\mathrm{eff}} = 1.0$, i.e., perfect sticking}
    \label{fig4}
\end{figure}

Dust coagulation will influence the ST. We calculate the disk torque on the planet using Equations (26)–-(28) in \cite{Hou2024}. To ensure both numerical convergence and computational efficiency, the torque density is integrated over $x \in [ -10H,\,10H ] $, uniform discretized into $100001$ grid points, and over $k_y \in [ 0.01 H^{-1}, \,15 H^{-1} ] $, logarithmically discretized into $320$ points.

Our results are presented in \autoref{fig4}, which shows the dusty torque on the planet with different values of $\varepsilon_{\mathrm{eff}}$, $\alpha$ and $\tilde{\mu} \equiv \mu / (1/99)$. We do not plot the gaseous torque because it is almost unaffected by dust coagulation. Lines with different colors correspond to different $\alpha$ as the legend labels. Dashed lines represent the results with $\varepsilon_{\mathrm{eff}} = 0$, i.e., without dust coagulation. Dot-dashed lines represent the results with $\varepsilon_{\mathrm{eff}} = 0.5$. Solid lines represent the results with $\varepsilon_{\mathrm{eff}} = 1.0$, i.e., perfect sticking.

With dust coagulation for $\varepsilon_{\mathrm{eff}} = 0.5$ or $1.0$, we can see that the dusty torque becomes more negative. For small $\tau$, the effect is negligible, while for large $\tau$, especially when $\tau \sim 0.5$, the effect is significant. But when $\tau$ is close to 1, dust coagulation becomes inefficient and imperfect sticking tends to occur. And an imperfect sticking makes the effect less significant. 

Comparing the results for different $\alpha$ values, we found that the effects of dust coagulation on the stopping time can be mitigated by dust diffusion. The influence of dust diffusion is easily understood as its smoothing effects. \cite{Hou2025} found that the ST is approximately linear with the dust mass fraction. We find that this linearity still holds when dust coagulation is included, as illustrated by the dotted orange line in \autoref{fig4}.

For a CM with a given $k_y$, we found that $\delta \Sigma_d$ leads $\delta \tau$ by $\pi/2$ in phase in \autoref{mechanism}. Then the integrated $\delta \Sigma_d$ over all $k_y$ will also leads the integrated $\delta \tau$ by $\pi/2$. \autoref{fig2} shows that the integrated $\delta \tau$ reaches its minimum along the streamline that crosses the planet. Then divided by the suppressed dust size wake, a density decrease will appear on the leading side, while a density enhancement will appear on the trailing side. This density asymmetry induced by phase shift leads to a net reduction in ST.

Up to this point, we note that although the dust coagulation timescale, $t_{\mathrm{coa}} = 3t_0/ \mu \simeq 150 \Omega_{p}^{-1}$, is relatively long, it can still exert a significant influence on ST. A natural concern is whether such a long timescale can really have any effect as dust grains pass by the planet. Here it is important to recognize that the CM is an advected mode, and the associated density variations evolve on the timescale of dust drift. For $\mu = 1/99$, $h_p = 0.05$, and $\tau = 0.1$, the dust drift time across one disk scale height is $t_{H} = |H/u_{dx}| \simeq 100 \Omega_p^{-1}$. Since $t_{\mathrm{coa}}$ and $t_{H}$ are comparable, the role of dust coagulation in modifying ST is physically well justified.

\subsection{The Effects of Imperfect Sticking and Fragmentation} \label{imperfect}
As shown in \autoref{fig3} and \autoref{fig4}, CMs-induced effects, the suppression of dust growth and streaming torque, are diminished when sticking is imperfect, which is physically intuitive. Mathematically, the amplitudes of CMs is shaped by both the planetary potential and $\Im(k_{x})$, implying that the influence of CMs on dust growth and ST scales positively with $\varepsilon_{\mathrm{eff}}$, consistent with the discussion in \autoref{mode}.

In the case of fragmentation, $\Im(k_{x})$ becomes negative, indicating that CMs have negligible effects. We confirm this by finding that CMs have minimal influence on ST. 

\subsection{Disk Torque Map} \label{torque_map}

\begin{figure}[htbp!]
    \centering
    \includegraphics[width=1.0\columnwidth]{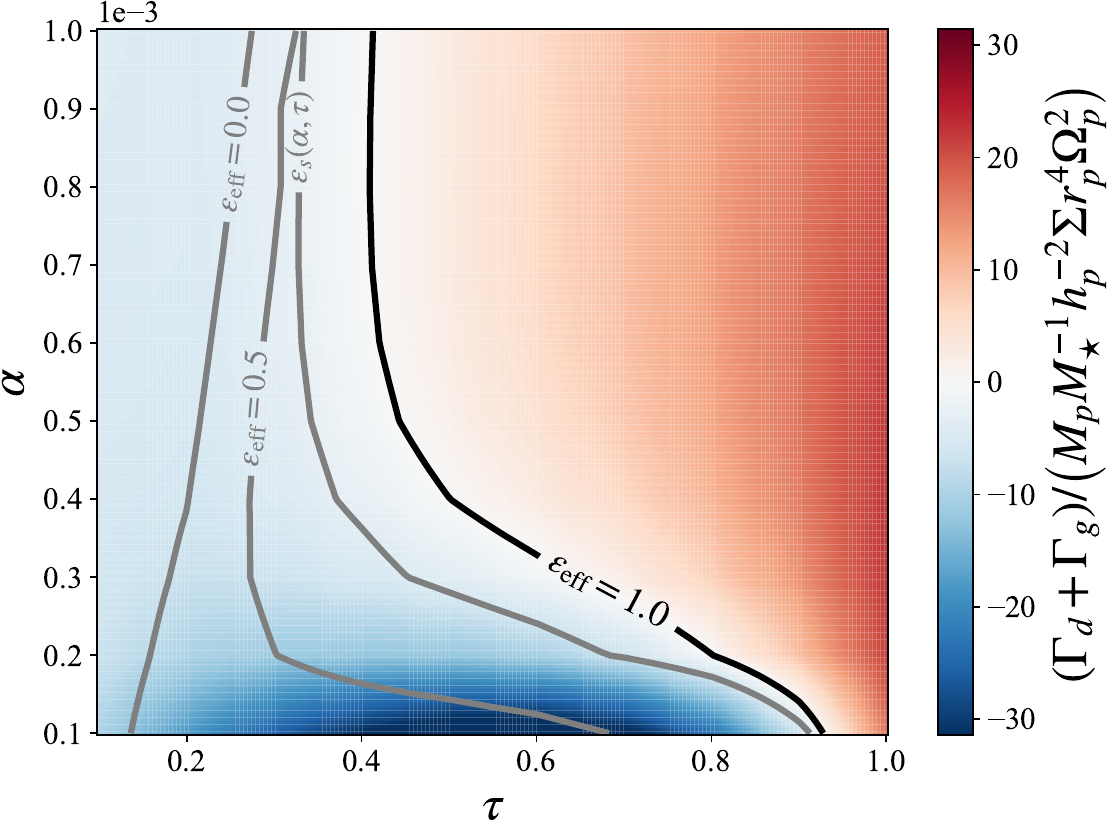}
    \caption{Disk torque map with perfect sticking. The $y$ axis represents gaseous turbulent strength. The black solid line marks the zero-torque contour, while three gray solid line shows the zero-torque contours with different sticking efficiency.}
    \label{fig7}
\end{figure}

Finally, we present the torque map on the planet, including contributions from both gas and dust, under the assumption of perfect sticking in \autoref{fig7}. The $\tau$ ranges from 0.1 to 1.0, and the vertical axis shows the gas turbulent strength $\alpha$, spanning $10^{-4}$ to $10^{-3}$. The black solid line marks the zero-torque contour. For comparison, three gray lines indicate zero-torque contours for $\varepsilon_{\mathrm{eff}} = 0$, 0.5, and $\varepsilon_s(\alpha,\tau)$, where $\varepsilon_s(\alpha,\tau)$ is computed using \autoref{eq_eff} with $v_{\mathrm{frag}} = 30\ \mathrm{m\ s^{-1}}$ and $c_s = 500\ \mathrm{m\ s^{-1}}$, typical values at $\sim 5\ \mathrm{AU}$ in PPDs.

In \autoref{fig7}, red regions indicate positive torque that drives outward migration, while blue regions correspond to inward migration. We find that the presence of CMs shifts the zero-torque contours toward larger $\tau$, thereby promoting inward migration. Under moderate turbulence and imperfect sticking, the critical $\tau$ for zero torque is approximately $0.3$, which will be a typical value in PPDs. In the future, more detailed modeling of dust growth can refine this estimate.

\section{Discussions and Conclusions} \label{conclusions}
In this study, we investigated the impact of low-mass planets on dust coagulation using a single-size approximation that incorporates CMs. We demonstrated the similarities between quasi-drift modes and CMs and revisited the mechanism of CI, highlighting the key role of phase shifts among perturbations in stopping time, dust density, and radial velocity. Our results show that although dust density remains high near the planet, the presence of CMs suppresses dust growth in the planets' co-orbital region. This suppression is less pronounced at smaller stopping times, stronger turbulence, and imperfect sticking.

We further examined how dust coagulation influences ST. We found that CMs reduce ST due to the $\pi/2$ phase shift between stopping time and dust density perturbations. For a low-mass planet to undergo outward migration, $\tau \gtrsim 0.3$ is required with typical turbulent strength and dust coagulation efficiency. In this study, the dust-to-gas ratio is mostly assumed to be $1/99$, and the ST is found to be approximately linear with this ratio.

Our findings have important implications for planetary formation and evolution. Planetary migration has long been recognized as a key process affecting planetary survival \cite[e.g.,][]{Ward1997,Masset2003,Yu2010,D'Angelo2010,Paardekooper2011,Tanaka2024,Wu2025,Chen2025}. Previous studies have shown that dusty torques (or ST) can drive outward migration \citep{Llambay2018,Guilera2023,Hou2024,Hou2025,Guilera2025,Chametla2025b}. Our results indicate that dust coagulation can significantly modify the dusty torque. Moreover, pebble accretion can enhance dusty torques \citep{Regaly2020,Chrenko2024}, but it relies on the efficient delivery of pebbles, i.e., dust particles in the intermediate $\tau$ range, marginally coupled to the gas with stopping times comparable to the orbital timescale, to planetary cores. This process may be affected by the suppressed dust size wake induced by CMs, as shown in our results. These findings point to a potentially important interplay between dust coagulation and pebble accretion, which could shape the dust-driven component of planetary migration. Understanding this coupling is essential for constructing comprehensive models of planet formation and migration.

While our model provides useful insights, it has several limitations. The suppression of dust growth by planets, as identified in this study, is primarily qualitative. And we are unable to quantify its long-term impact. Further calculations are required to capture the cumulative effects over longer timescales. In addition, nonlinear effects, particularly those associated with the horseshoe region, are not included and may introduce additional complexities in dust dynamics. We note that our results rely on the single-size approximation, which assumes that the dust mass distribution is narrowly peaked at a representative mass and that collisions between similar-size particles dominate mass growth. While the validity of this approach has been confirmed by \cite{Sato2016}, it neglects unequal-mass collisions, fragmentation, and the detailed evolution of the full size distribution. Future work solving the full Smoluchowski equation will be important to fully capture these processes and more accurately assess dust evolution and planetary influence. Regarding the torque calculations, they are based on a 2D shearing-sheet model, which is appropriate for evaluating the ST \citep{Hou2025}. Although 3D effects are known to modify gas torques, their influence on the dusty torque remains uncertain \citep{Chametla2025a}, including the dependence on the adopted smoothing length. Further studies incorporating full 3D effects are needed.

\section*{Acknowledgment}
We thank the anonymous referee for the helpful comments and suggestions that clarified and improved the manuscript. Q.H. thank Ryosuke T. Tominaga for helpful discussions. This work is supported by the National Natural Science Foundation of China (NSFC, No. 12288102). C.Y. has been supported by the National SKA Program of China (Grant No. 2022SKA0120101) and National Key R\&D Program of China (No. 2020YFC2201200) and the science research grants from the China Manned Space Project (No. CMS-CSST-2021-B09, CMS-CSST-2021-B12, CMSCSST-2021-A10, and CMS-CSST-2025-A16) and opening fund of State Key Laboratory of Lunar and Planetary Sciences (Macau University of Science and Technology) (Macau FDCT Grant No. SKL-LPS(MUST)-2021-2023) and National Natural Science Foundation of China (grants 11521303, 11733010, 11873103, and 12373071).

\appendix 
\section{Linear Perturbation Equations} \label{appendix}
The linear perturbation form of \autoref{eq1}-\ref{eq4} and \autoref{eq5} reads
\begin{gather}
    \nabla_g \frac{\delta \Sigma_g} {\Sigma_g}+ \frac{d}{d x} \delta u_{gx} + i k_y \delta u_{gy} = 0, \label{linear1}\\
    \frac{d}{dx} \frac{\delta \Sigma_g}{\Sigma_g} + \nabla_g \delta u_{gx} - 2 \Omega_p \delta u_{gy} - \frac{ w_{s,x}}{t_s} \frac{\delta \Sigma_d}{\Sigma_g} - \frac{\mu }{t_s} \delta w_{s,x} + \frac{\mu w_{s,x}}{t_s} \frac{\delta t_s}{t_s} = - \frac{\partial \phi_p}{\partial x}, \label{linear2} \\ 
    i k_y \frac{\delta \Sigma_g}{\Sigma_g} + \frac{\Omega_p}{2}\delta u_{gx} + \nabla_g \delta u_{gy} -  \frac{w_{s,y}}{t_s}\frac{\delta \Sigma_d}{\Sigma_g}  - \frac{\mu }{t_s} \delta w_{s,y} + \frac{\mu w_{s,y}}{t_s} \frac{\delta t_s}{t_s} = - i k_y \phi_p, \label{linear3} \\
    \nabla_d \frac{\delta \Sigma_d}{\Sigma_d} + \frac{d}{d x} \delta u_{dx} + i k_y \delta u_{dy} = 0, \label{linear4}\\
    \nabla_d \delta u_{dx} - 2 \Omega_p \delta u_{dy} + \frac{1}{t_s} \delta w_{x} + \frac{ w_{s,x}}{t_s} \frac{\delta \Sigma_g}{\Sigma_g} - \frac{w_{s,x}}{t_s} \frac{\delta t_s}{t_s} + \frac{\alpha}{t_s} \left[ \frac{d }{dx} \left(\frac{\delta \Sigma_d}{\Sigma_d}\right) - \frac{d }{dx} \left(\frac{\delta \Sigma_g}{\Sigma_g}\right) \right] = - \frac{\partial \phi_p}{\partial x}, \label{linear5}\\
    \frac{\Omega_p }{2} \delta u_{dx} + \nabla_d \delta u_{dy} + \frac{1}{t_s} \delta w_{s,y} + \frac{ w_{s,y}}{t_s} \frac{\delta \Sigma_g}{\Sigma_g} - \frac{ w_{s,y}}{t_s} \frac{\delta t_s}{t_s} + \frac{i k_y \alpha}{t_s} \left( \frac{\delta \Sigma_d}{\Sigma_d} - \frac{\delta \Sigma_g}{\Sigma_g} \right)= - ik_y \phi_p, \label{linear6} \\
    \nabla_d \frac{\delta t_s}{t_s} - \frac{\varepsilon_{\mathrm{eff}}}{3 \Sigma_g t_0} \left( \delta \Sigma_d - \mu \delta \Sigma_g +\Sigma_d \delta t_s \right) - \frac{d}{dx} \delta u_{gx} - i k_y \delta u_{gy} + \left( i k_y w_{s,y} + w_{s,x} \right) \frac{\delta \Sigma_g}{\Sigma_g} = 0, \label{linear_coa}
\end{gather}
where $\delta w_{s,x} \equiv \delta u_{dx} - \delta u_{gx}$, $\delta w_{s,y} \equiv \delta u_{dy} - \delta u_{gy}$, $\nabla_g \equiv i k_y u_{gy} + u_{gx} \frac{d}{d x}$ and $\nabla_d \equiv i k_y u_{dy} + u_{dx} \frac{d}{d x}$. The planetary potential in Fourier space, $\phi_p$,  is expressed as the Bessel functions \citep[e.g., Equations (19)-(20) in][]{Hou2024}. 
\bibliography{ST_coa.bib}{}
\bibliographystyle{aasjournal}
\end{document}